# Comment on "Electromagnetic Surface Modes at Interfaces with Negative Refractive Index make a Non-Quite Perfect Lens"

N. Garcia (1) and M. Nieto-Vesperinas (2)

(1) E-address: nicolas.garcia@fsp.csic.es, Laboratorio de Fisica de sistemas Pequeños y Nanotecnología, Consejo superior de Investigaciones Cientificas, Serrano 144, Madrid 28006, Spain.

(2) E-address: mnieto@icmm.csic.es, Instituto de Ciencia de Materiales de Madrid, Consejo Superior de Investigaciones Cientificas, Campus de Cantoblanco, Madrid 28049, Spain

In [1] a discussion and plot of surface modes in a slab of left-handed material (LHM), is made. We wish to point out that, except these facts, that work does not add anything that has not already been said or implitely contained in the transmittivity and reflectivity equations of [2]. However, the main point in [1] is the existence of a cut-off in the wavevector, $k_c = 2\pi/a$ as a consequence of the failure of the effective, continuous, medium theory based on Maxwell's equations, and the subsequent existence of Bragg diffraction by the inner periodic structure of the medium, of period $a$. Such cut-off then prevents the non-integrability of the field wavefunction inside the LHM pointed out in [2], and named ultraviolet divergence in [1].

For metamaterials, $a = 0.5 cm$, therefore, this size has nothing to do with superresolution. Thus, there is no point in discussing it. However, assuming a crystal with $n = -1$, or for example, the case of silver addressed in [3], $a$ is now the atomic lattice period, hence, $k_c = 2\pi Å^{-1}$.

In [1] it is stated that in [2] the incorrect conclusion was drawn that the above mentioned divergence would prevent the electromagnetic wave to penetrate inside the slab. However, in [2] it was very clearly stated that this lack of transmission into the LHM was due to the unlimited time required to build up such a surface state, and thus to the infinite energy involved, as shown next. However, [1] simply addresses the cut-off without caring about this problem, namely, the number of photons involved in filling that state. This is, however, the physics one has to care about of. Let us discuss this point as regards the superlens proposed in [3].

Let the distance between the light source and the slab be $20 nm$, and the slab thickness $40 nm$. Then the intensity amplifying factor at the exit interface of the slab is: $\exp(4\pi \times 200) \simeq 10^{1091}$. Now, the numer of photons in the plate $n\hbar\omega$ equals $\frac{V}{8\pi}|E|^2 A^2$ where $E$ is the electric field amplitude and $A^2$ stands for its aforementioned amplification factor as it traverses the slab. On the other hand, $V$ represents the plate volume iluminated by the passing wave.

Let us assume that one wants to retrieve only one photon per year at the other side of the slab, (even if little, one should detect something!). From the above, this requires a source providing $10^{1081}$ photons per second!. What kind of source is this?, we believe that this is simply nonsense. Not to mention what is the material capable of enduring such energy. Conversely, it can be argued that one will do all this superlensing to recover just $10^{-1091}$ photons per year!. At the light frequencies, which is were the superlensisng was proposed, our analysis shows no chances for superresolution, in that region, near field optical micrsocopies are more effective A different matter is at microwave frequencies. This however requires further study, and in any case it has nothing to do with nanometer resolution as discussed in the configuration of [3].

Ref. 1 also argues that no losses need to be considered in the medium with negative refractive index. This is not true, because as shown [6], the medium must be dispersive, and this in turn implies losses. Then, no matter how small they are, their effect will appear before that of the cut-off. In this connection, we recall two recent papers [5] and [4]

In conclusion as we said in ref. 2 (2nd page column 2, lines 10-13) the wave cannot go through the LHM because the building transient time to conform the wave in the medium will be infinite, so that the energy accumulated in the slab will be so large that it will destroy the material. Even better than any lens, the material will be a fantastic energy reservoir!

[1] F.D.M. Haldane, cond-mat/0206420.
[2] N. Garcia and M. Nieto-Vesperinas, Phys. Rev. Lett. **88**, 207403 (2002).
[3] J.B. Pendry, Phys. Rev. Lett. **85**, 3966 (2000).
[4] N.Garcia and M. Nieto-Vesperinas, COND-MAT/0207320 and COND-MAT/0206460
[5] J. T. Shen and P. M. Platzman, Appl. Phys. Lett. 80, 3286 (2002).
[6] L. D. Landau and E.M. Lifshitz, Electrodinámica de los

Medios Continuos (Editorial Reverté, Barcelona España 1981).